\documentclass[conference]{IEEEtran}
\usepackage[T1]{fontenc}
\usepackage[utf8]{inputenc}
\usepackage{cite}
\usepackage{amsmath,amssymb}
\usepackage{url}
\usepackage{hyperref}
\hypersetup{hidelinks}

\title{Smart Contract Security Beyond Detection}

\author{\IEEEauthorblockN{Tamer Abdelaziz, Ph.D.}
\IEEEauthorblockA{tamer.m@nyu.edu}}

\begin{document}

\maketitle

\begin{abstract}
Smart contract security has progressed from vulnerability detection toward a broader research agenda that includes semantic reasoning, automated repair, adversarial robustness, and real-time exploit detection. This paper develops a capstone-oriented research narrative around four directions: foundation-model-based smart contract semantics and vulnerability reasoning~\cite{abdelaziz2025foundation}, automated smart contract repair with formal guarantees~\cite{abdelaziz2025automated}, adversarial learning for robust malicious contract and transaction detection~\cite{abdelaziz2025adversarial}, and real-time transaction-level exploit detection at blockchain scale~\cite{abdelaziz2025real}. We connect these directions to two recent studies that characterize the current frontier: a diagnostic analysis of where smart contract security analyzers fall short~\cite{abdelaziz2026smart} and a scalable real-time system for malicious Ethereum transaction detection~\cite{abdelaziz2026txlens}. The resulting framework is intended to help students formulate capstone projects that are technically grounded, empirically measurable, and aligned with contemporary smart contract security research.
\end{abstract}


\section{Introduction}

Smart contracts have transformed blockchain platforms into engines for decentralized finance, digital assets, and automated agreement execution~\cite{buterin2013ethereum}. By enabling trustless, self-executing code on blockchain networks, they have supported rapid innovation across decentralized finance (DeFi) protocols~\cite{werner2022sok} and non-fungible token (NFT) ecosystems~\cite{wang2021non}. At the same time, the blockchain industry continues to expand rapidly: the global blockchain market is projected to grow from USD 32.99 billion in 2025 to USD 393.45 billion by 2030, at a compound annual growth rate (CAGR) of 64.2\%~\cite{marketsandmarkets_blockchain_2025}. This growth reflects the increasing reliance on blockchain infrastructure across finance, logistics, digital identity, and other security-sensitive domains.

The same properties that make smart contracts attractive also make them difficult to secure. Immutability means that once a contract is deployed, vulnerabilities may persist indefinitely unless an upgrade or migration mechanism exists. Permissionless composability increases the attack surface by allowing contracts to interact in complex and often unpredictable ways. These properties have made security failures especially costly. CertiK's \textit{Hack3D: The Web3 Security Report 2025} reports 630 security incidents in 2025, resulting in approximately USD 3.35 billion in losses, a 37\% increase from the previous year; Ethereum alone accounted for 310 hacks, scams, and exploits, with USD 1,697,833,313 in losses~\cite{certik_hack3d_2025}. These figures underscore that smart contract security is not a niche concern but a major systems and economic challenge.

Research on smart contract security has therefore expanded beyond vulnerability identification to include richer forms of program understanding, repair, and deployment. The central question is no longer only whether a vulnerable contract can be detected, but whether security tools can reason about semantics, synthesize safe fixes, resist adversarial manipulation, and operate under real-time constraints. This broader perspective is reflected in four recent research directions: foundation models for semantic understanding and vulnerability reasoning~\cite{abdelaziz2025foundation}, automated repair with formal guarantees~\cite{abdelaziz2025automated}, adversarial learning for robust malicious contract and transaction detection~\cite{abdelaziz2025adversarial}, and real-time exploit detection at blockchain scale~\cite{abdelaziz2025real}.

Two additional studies sharpen this agenda. One examines where current analyzers fail in practice and exposes structural blind spots that remain unresolved~\cite{abdelaziz2026smart}. The other demonstrates the importance of scalable transaction-level monitoring for malicious activity detection in live Ethereum traffic~\cite{abdelaziz2026txlens}. Taken together, these works suggest a coherent progression from understanding to detection to defense. This paper turns that progression into a structured capstone framework that students can use to identify a meaningful, technically defensible project topic.

\section{Related Work}

\subsection{Semantic reasoning for smart contract understanding}
The foundation-model direction argues that smart contract analysis should go beyond token-level similarity and local syntactic cues. Instead, models should reason over code structure, control flow, data flow, and state transitions in order to capture security-relevant behavior~\cite{abdelaziz2025foundation}. A meaningful interpretation of this direction is that a model should infer whether a contract's observable behavior aligns with its intended security properties. This requires representations that account for function interaction, access control, external calls, and execution context. For capstone work, this opens several directions, including semantic embeddings for contract functions, explanation-aware vulnerability prediction, and reasoning-oriented classifiers that better capture long-range dependencies.

\subsection{Automated repair with formal constraints}
The repair direction moves beyond classification by asking whether a vulnerable contract can be transformed into a safer version without compromising intended functionality~\cite{abdelaziz2025automated}. This requires a synthesis pipeline that localizes faults, generates candidate patches, and validates them against correctness criteria. In a strong formulation, repair should satisfy invariants such as type consistency, control-flow preservation, and behavioral equivalence on relevant traces. The most compelling capstone projects in this area combine patch generation with symbolic checking, invariant reasoning, or test-guided validation so that a patch is not only syntactically valid, but also semantically justified.

\subsection{Adversarial robustness of learned detectors}
The adversarial-learning direction highlights a persistent weakness of machine learning in security: a model that performs well on clean data may degrade sharply under adversarial perturbation~\cite{abdelaziz2025adversarial}. In smart contract security, such perturbations may include code obfuscation, insertion of semantically neutral statements, identifier renaming, reordering of independent operations, or transaction-level manipulations intended to evade detection. A technically sound capstone project in this area should define a threat model, measure sensitivity to perturbations, and evaluate whether adversarial training, augmentation, or robust representations improve resistance. The key question is not only whether accuracy improves, but whether the detector remains stable when confronted with an attacker actively trying to conceal malicious behavior.

\subsection{Real-time exploit detection at blockchain scale}
The real-time detection direction reframes smart contract security as an operational monitoring problem rather than solely an offline analysis task~\cite{abdelaziz2025real}. In this setting, a detector must process transaction streams quickly enough to provide actionable alerts before damage propagates. This introduces constraints on inference latency, memory usage, batching strategy, and feature availability at decision time. A capstone in this direction may therefore focus on streaming feature extraction, low-latency classification, threshold calibration, or alert prioritization. The core challenge is to preserve detection quality while meeting strict runtime requirements, which is substantially more difficult than achieving high offline accuracy.

\subsection{Diagnostic analysis of analyzer shortcomings}
The analyzer-failure study is especially valuable because it frames smart contract security as an evaluation problem as well as a detection problem~\cite{abdelaziz2026smart}. A mature interpretation is that security tools should be assessed not only by aggregate metrics, but also by the types of vulnerabilities they miss, the contract structures that trigger failures, and the code patterns that produce false positives. This motivates capstone projects focused on systematic error analysis, failure taxonomy construction, and comparative evaluation across analyzers. Such work can produce important insights even without proposing a new detection model, particularly when it explains why existing approaches remain incomplete.

\subsection{Scalable transaction-level malicious activity detection}
TxLens extends the deployment-oriented direction by emphasizing scalable, real-time detection of malicious Ethereum transactions~\cite{abdelaziz2026txlens}. From a technical standpoint, this implies high-throughput streaming analysis, efficient feature computation, and a model architecture that can operate under tight latency budgets. A capstone inspired by this work could explore online inference pipelines, detector calibration for live traffic, or the trade-off between sensitivity and throughput. The central insight is that blockchain defense must function in a streaming environment where decisions are time-critical.

\section{Methodology}

A capstone project in this area should follow a reproducible and technically disciplined methodology.

\subsection{Problem definition}
The student should select one primary objective: semantic reasoning, repair synthesis, adversarial robustness, or real-time detection. The project should state a precise research question and define success criteria. For example, a semantic reasoning project may ask whether richer representations improve vulnerability discrimination; a repair project may ask whether synthesized patches preserve intended behavior; an adversarial project may ask whether performance degrades under obfuscation; and a real-time project may ask whether the system satisfies a target latency bound.

\subsection{System design}
The implementation should be decomposed into clearly documented stages. A learning-based project should define preprocessing, representation learning, model training, and evaluation. A repair-based project should define bug localization, patch generation, and validation. A streaming project should define event ingestion, feature extraction, inference, and alerting. In all cases, the pipeline should support reproducibility, ablation studies, and clear error analysis.

\subsection{Evaluation protocol}
A strong evaluation should include both effectiveness and robustness. Detection tasks should report precision, recall, F1, false-positive rate, and confusion matrices. Robustness studies should include perturbation-based evaluation or adversarial stress tests. Repair tasks should report patch correctness, compilation success, and semantic preservation checks. Real-time systems should report throughput, latency, and memory usage. Whenever feasible, the project should compare multiple baselines and analyze failure cases to explain where the method succeeds or fails.

\subsection{Contribution criteria}
To be capstone-worthy or publishable, the project should contribute at least one of the following: a new semantic representation, a repair method with validation, a robustness mechanism, an error taxonomy, or a deployment-oriented optimization. The novelty may be incremental, but it should be concrete, measurable, and clearly positioned relative to existing methods.

\section{Capstone Ideas}

\subsection{Semantic reasoning with foundation models}
This project would investigate whether foundation-model representations improve the understanding of smart contract behavior relative to conventional token-based or static-feature methods~\cite{abdelaziz2025foundation}. The student could encode entire functions, control-flow fragments, or contract summaries and evaluate whether the model better captures security-relevant semantics. A stronger version of the project would also generate explanations that identify which contract components influence the prediction. The goal is to determine whether semantic representation learning can reduce reliance on shallow syntactic cues.

\subsection{Automated repair with correctness validation}
This project would generate candidate fixes for vulnerable contracts and then validate whether the repairs are semantically safe~\cite{abdelaziz2025automated}. The main technical challenge is to ensure that a repair preserves intended functionality while eliminating the vulnerability. The student could combine patch generation with symbolic checking, invariant checking, or test-based validation. A strong outcome would be a repair pipeline that proposes multiple candidate patches, ranks them by plausibility, and rejects those that violate correctness constraints.

\subsection{Adversarially robust detection}
This project would study how smart contract detectors behave under code perturbations and evasion-oriented transformations~\cite{abdelaziz2025adversarial}. The student should define a threat model and test whether adversarial training or augmentation improves resilience. The project could measure degradation under renaming, statement insertion, function reordering, or transaction-level manipulation. A strong contribution would be a detector whose robustness curve is substantially flatter than that of a standard baseline.

\subsection{Real-time exploit monitoring}
This project would focus on streaming detection of malicious blockchain activity under realistic deployment constraints~\cite{abdelaziz2025real,abdelaziz2026txlens}. The technical challenge is to build a pipeline that ingests transactions, computes features quickly, and produces alerts with bounded latency. The student could study batching strategies, threshold calibration, and the trade-off between recall and throughput. This direction is especially suitable for capstone projects that bridge machine learning and systems engineering.

\subsection{Analyzer failure analysis}
This project would compare existing analyzers and study where they fail most often~\cite{abdelaziz2026smart}. The objective is to construct a taxonomy of missed vulnerabilities, false positives, and structural blind spots. The student could cluster failure cases by vulnerability type, contract structure, or control-flow pattern and then explain which analyzer assumptions are violated. Such a study is valuable because it identifies which problems remain unresolved even when aggregate scores appear strong.

\section{Discussion}

The central message of these works is that smart contract security has matured into a multi-stage research problem. Detection remains important, but it is no longer sufficient. A useful security tool must understand semantics, resist manipulation, support repair, and operate at blockchain speed.

This creates a strong environment for capstone work because students can contribute at different technical levels. A project does not need to introduce a new theory of security to be valuable. It can improve semantic modeling, strengthen robustness, validate repairs, or analyze system failures. The critical criterion is that the work be technically explicit, experimentally grounded, and clearly tied to a real security need.

The recent diagnostic and systems papers are especially useful because they reveal both the limitations of current analyzers and the requirements of practical deployment~\cite{abdelaziz2026smart,abdelaziz2026txlens}. Together with the four proposed directions, they provide a coherent research path from understanding to detection~\cite{mohamed2023towards,abdelaziz23usenix} to defense.

\section{Conclusion}

This paper presented a capstone-oriented research narrative for smart contract security based on four technical directions: semantic reasoning, formal repair, adversarial robustness, and real-time exploit detection~\cite{abdelaziz2025foundation,abdelaziz2025automated,abdelaziz2025adversarial,abdelaziz2025real}. We then connected these directions to recent work on analyzer shortcomings and scalable malicious transaction detection~\cite{abdelaziz2026smart,abdelaziz2026txlens}. The resulting framework offers students a professional academic basis for selecting a topic, identifying its technical requirements, and positioning its relevance within the current smart contract security literature.

\bibliographystyle{IEEEtran}
\bibliography{references}

\end{document}